\documentclass[aps,prl,showpacs,twocolumn,preprintnumbers,superscriptaddress]{revtex4}
\usepackage{amsmath,amssymb,graphicx,color,bm,epstopdf}

\begin{document}

\title{Semiconductor cavity QED: Bandgap induced by vacuum fluctuations}

\author{T. Espinosa-Ortega}
\affiliation{Division of Physics and Applied Physics, Nanyang
Technological University 637371, Singapore}

\author{O. Kyriienko}
\affiliation{Division of Physics and Applied Physics, Nanyang
Technological University 637371, Singapore} \affiliation{Science
Institute, University of Iceland, Dunhagi-3, IS-107, Reykjavik,
Iceland}

\author{O. V. Kibis}
\affiliation{Division of Physics and Applied Physics, Nanyang
Technological University 637371, Singapore}
\affiliation{Department of Applied and Theoretical Physics,
Novosibirsk State Technical University, Karl Marx Avenue 20,
630073 Novosibirsk, Russia}

\author{I. A. Shelykh}
\affiliation{Division of Physics and Applied Physics, Nanyang
Technological University 637371, Singapore} \affiliation{Science
Institute, University of Iceland, Dunhagi-3, IS-107, Reykjavik,
Iceland}


\begin{abstract}
We consider theoretically a semiconductor nanostructure embedded in one-dimensional microcavity and study the modification of its electron energy spectrum by the vacuum fluctuations of the electromagnetic field. To solve the problem, a non-perturbative diagrammatic approach based on the Green's function formalism is developed. It is shown that the interaction of the system with the vacuum fluctuations of the optical cavity opens gaps within the valence band of the semiconductor. The approach is verified for the case of large photon occupation numbers, proving the validity
of the model by comparing to previous studies of the semiconductor system excited by a classical electromagnetic field. The developed theory is of general character and allows for unification of quantum and classical descriptions of the strong light-matter interaction in semiconductor structures.
\end{abstract}


\maketitle

\textit{Introduction.---}The interaction between light and matter is an important part of modern physics, interesting from both a fundamental and an applied point of view. The investigation of the regime of strong light-matter coupling, where the interaction between photons and material excitations can not be treated perturbatively, is of special interest. One of the fundamental phenomena in this domain is the dynamic (AC) Stark effect \cite{Cohen-Tannoudji} associated with the stationary energy shift of electron energy levels under the influence of an electromagnetic wave and taking place in both atomic systems and solids \cite{Koch,Faist,Hayat_12,Vidar,Galitskii,Goreslavskii,Vu2004,Kibis_PRL2011,Kibis_PRB2012}. Particularly, the dynamic Stark effect opens stationary bandgaps in semiconductor systems, which take place in resonant points of the Brillouin zone satisfying the condition where the photon energy is equal to the energy interval between electron bands of the semiconductor. This gap can manifest itself in various physical effects \cite{Vu2004,Kibis_PRL2011,Kibis_PRB2012}.

Phenomena similar to the dynamic Stark effect can take place not only for electrons interacting with strong classical electromagnetic waves, but for electrons interacting with vacuum fluctuations
of the electromagnetic field as well. In the latter case, the accounting for the corrections given by quantum electrodynamics (QED) becomes crucial. Their study has begun with experimental observation and theoretical explanation of the Lamb shift \cite{Lamb_47,Bethe_47} and the Casimir effect \cite{Casimir,Dutra}. In the domain of cavity QED, vacuum field fluctuations were shown to induce vacuum Rabi oscillations which leads to the shift of atomic energy levels \cite{Vahala2003}, the generation of correlated photonic pairs \cite{Ciuti}, and persistent currents in a quantum ring \cite{Kibis_PRB2013}.

Though the dynamic Stark effect has been known longer than half a century, for solids the previous studies were either focused on the case of strong (``classical'') electromagnetic wave, where the quasiclassical description of the field is valid (see, \textit{e. g.}, Refs. \cite{Galitskii,Goreslavskii}), or the electron interaction with photons can be treated using a perturbation theory \cite{Kibis_PRB2013,Kibis_graphene}. As to a consistent non-perturbative quantum theory of the effect, it was unknown before. In order to fill this gap in the theory, in the given paper we develop a non-perturbative approach which is applied to study the dynamic Stark effect in one-dimensional (1D) cavities. Conceptually, the approach is based on summation of an infinite series of Feynman diagrams describing emissions and absorptions of cavity photons \cite{Bruus}. Using the Dyson equation for renormalized electron Green's functions, we calculate the electron dispersion and reveal the appearance of photon-induced bandgaps in the density of electron states. We show that our theory reconfirms previous results in the classical limit of strong electromagnetic wave \cite{Galitskii,Goreslavskii} and allows for correct calculation of the vacuum-induced bandgaps. Such a merging of the concepts of vacuum-induced corrections to electron energy spectra and strong light-matter coupling in microcavities opens an interdisciplinary area of research, lying in the boundary between quantum electrodynamics and condensed matter physics.

\begin{figure}
\includegraphics[width=0.8\linewidth]{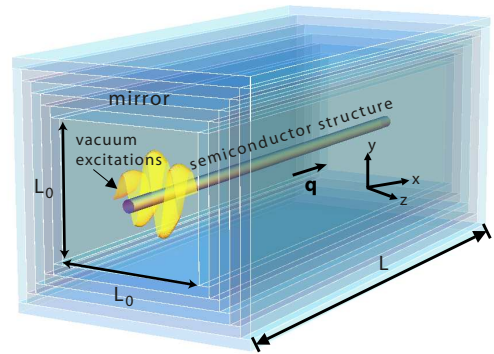}
\caption{(Color online) Sketch of the system. The one-dimensional cavity with a semiconductor structure embedded in the antinode. Since the cavity dimensions satisfy the condition $L\gg L_0$, the cavity photon mode is confined along the $y,z$ axes and can propagate with the wave vector $\mathbf{q}=(q,0,0)$ only along the $x$ axis. The same is assumed to be true for electrons and holes in the structure.} \label{Fig1}
\end{figure}

\textit{Model.---}We analyze a semiconductor quantum wire embedded in the antinode of a cavity, as shown in Fig.~\ref{Fig1}.
Let us consider the physically relevant case where the valence band is filled by electrons and the conduction band is initially empty. We account for a single waveguiding mode of the 1D cavity and a single subband in the conduction and valence band of the wire. These assumptions allow us to simplify the consideration and get analytical results. The generalization for the multimode case is straightforward.

Stimulated by vacuum electromagnetic fluctuations, an electron in the valence band (with energy
$E_v(k)$ and wave vector $k$ along the wire) can absorb a vacuum photon with energy $\nu$ and wave vector
$q$, and make a transition to an empty state in the conduction band with energy $E_c(k+q)$.
Consequently, it can emit the same photon and return to the initial state in the valence band. The process we described is called resonant (see the discussion of the Jaynes-Cummings model in Ref. \cite{Shore}), and it makes the main contribution to the electron-photon interaction in the case of classically strong electromagnetic field
\cite{Galitskii,Goreslavskii,Kibis_PRL2011}. However, in the case of vacuum fluctuations, we have to take into account also other---antiresonant---processes in which an electron, being initially in the valence band, gets excited to the conduction band with energy $E_c(k-q)$, and simultaneously emits a vacuum photon with energy $\nu$ and wave vector $q$.
Later the electron returns to its initial state in the valence band, with the absorption of the same photon. The aforementioned processes can take place many times, leading to hybridization of the electron and photon states that are typically referred to as electron state ``dressed'' by vacuum fluctuations. The anti-resonant processes can play a significant role in a variety of physical situations. In particular, they lead to the transition to ultra-strong \cite{Ciuti,Niemczyk2010,Todorov2010} and deep strong \cite{DeLiberato2014} coupling regimes, quantum phase transition from a normal to superradiant state in the Dicke model \cite{Hepp1973,Wang1973,Baumann2010,Nagy2010}, dynamical Casimir effect \cite{Ciuti,DeLiberato2007} and others.

In what follows, we will show that these vacuum-induced interband electron transitions can lead to the intraband gap opening in the same way as the transitions induced by a classical field \cite{Galitskii,Goreslavskii,Kibis_PRL2011}. We specially stress that the one-dimensional geometry of the photonic cavity is crucial for the effect. Physically, this is a consequence of the van Hove singularity in the photonic density of states at the point $q=0$. Indeed, the photonic dispersion can be estimated as $E(q)=c\hbar\sqrt{q^2+2\pi^2/L_0^2}$, where $L_0$ is the lateral size of the cavity, and thus the photonic density of states diverges as $\sim 1/\sqrt{q}$ at $q\rightarrow0$. As a result, vertical transitions with virtual photons become predominant, and the bandgap is opened as discussed below. We have checked that the effect is absent in planar microcavities, where the photonic density of states is regular.

\begin{figure}
\includegraphics[width=1.0\linewidth]{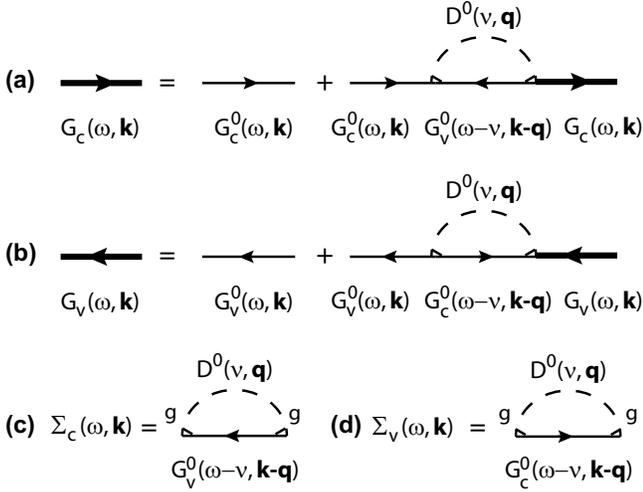}
\caption{The Dyson equations for renormalized electron Green's
functions corresponding to the conduction band (a) and the valence
band (b). The self-energy operators responsible for the
light-induced dressing of the conduction band (c) and the valence
band (d). The dashed line corresponds to a virtual cavity photon,
and $g$ denotes the electron-photon coupling constant.}
\label{Fig2}
\end{figure}

Formally, the accounting for all possible repetitions of resonant
and antiresonant vacuum-induced transitions can be done using the
Feynman diagrams which describe multiple absorptions and
re-emissions of vacuum photons (see Fig.~\ref{Fig2}). The process
corresponding to an electron being in the conduction band dressed
by the vacuum fluctuations can be represented in a diagrammatic
form by the series given in Fig.~\ref{Fig2}(a). The analogous case
for a dressed electron being initially in the valence band is
presented in Fig.~2(b). Using the diagrammatic representation, it
is possible to account for electron-photon interaction up to infinite order by solving the corresponding Dyson equation for the Green's function of the electrons dressed by electromagnetic field.

Realizing the aforementioned theoretical scheme, we restrict our
consideration to the valence band, but the proper
generalization for the conduction band can be easily made. 
Generally, a dressed electron in the valence band with energy
$\varepsilon$ and wave vector $k$ can be described by the
Dyson equation shown in Fig. \ref{Fig2}(b), which can be written
in the algebraic form as
\begin{equation}
\label{Greenf}
G_v(\varepsilon,k)=\frac{G_v^0(\varepsilon,k)}{1-\Sigma_v(\varepsilon,k)
G_v^0(\varepsilon,k)},
\end{equation}
where $G_v^0(\varepsilon,k)$ is the bare Green's function
for the electron in the valence band and
$\Sigma_v(\varepsilon,k)$ is the self-energy operator
depicted in Fig. \ref{Fig2}(d). The self-energy accounts for
interband transitions, where a photon with energy $\nu$ and wave
vector $q$ is absorbed and emitted (or conversely). It is
given by
\begin{equation}
\label{Sigma}
i\Sigma_v(\varepsilon,k)=-g^{2}L\int\limits_{\nu}\int\limits_{q}\frac{d\nu
dq}{(2\pi)^2}D^0(\nu,q)G_c^0(\varepsilon-\nu,k-q),
\end{equation}
where $G_c^0(\varepsilon,k)$ is the bare conduction
electron Green's function, $D^0(\nu,q)$ is the bare
photon Green's function, and the integration is performed over all
photon energies $\nu$ and photon wave vectors $q$. Here
the matrix element of the operator of the electron-photon interaction,
being the electron-photon coupling constant, reads
$g=|d_{cv}|\sqrt{{\omega_0(0)}/{2 \epsilon \epsilon_{0}SL}}$
\cite{Haug}, where $d_{cv}$ is the dipole matrix element of
interband transition, $\omega_0({q})$ is the energy of the cavity
photon, $\epsilon$ is the material permittivity, $\epsilon_0$ is
the vacuum permittivity, $S=L_0^2$ is the cross-section area of
the cavity, and $L$ is the cavity length. The expressions for the
bare Green's functions have the form
\begin{align}
\label{DO}
&D^0(\nu,q)= \frac{2\omega_0(q)}{(\nu-\omega_0(q)+i\delta)(\nu+\omega_0(q)-i\delta)},\\
\label{G0v}
&G_v^0(\varepsilon,k)=(\varepsilon-E_v(k)+i\delta)^{-1},\\
\label{G0c}
&G_c^0(\varepsilon,k)=(\varepsilon-E_c(k)-i\delta)^{-1},
\end{align}
where $\delta$ is the infinitesimally small shift arising from the
causality principle, the electron energy spectrum for the
conduction band and the valence band are
$E_{c}(k)= \hbar^2 k^2/2m_e + E_g/2$ and
$E_v(k)=-\hbar^2 k^2/2m_e - E_g/2$, the semiconductor
bandgap is $E_g$, and $m_e$ is the effective electron mass in the
semiconductor. The energy spectrum of the principal cavity photon
mode, $\omega_0({q})$, is given by
\begin{equation}
\label{oq} \omega_0({q})=\frac{\hbar
c}{n_0}\sqrt{q^2+q_y^2+q_z^2}\approx \omega_0 + \frac{\hbar^2
q^2}{2m_0},
\end{equation}
where $n_0$ is the refractive index of the medium,
$q_y=q_z=\pi/L_0$ are the quantized components of the photon wave
vector in the one-dimensional cavity, $\omega_0=\hbar c \pi
\sqrt{2}/n_0 L_0$ is the cavity photon rest energy, and $m_0=\hbar
n_0\pi/\sqrt{2}L_0c$ is the cavity photon effective mass.

It should be stressed that the dashed lines representing the
photon propagators have no preferable direction (see diagrams
shown in Fig.~\ref{Fig2}). This comes from the fact that the
photon Green's function $D^0(\nu,q)$ has two poles in
Eq.~(\ref{DO}). As a consequence, it leads to the simultaneous
accounting of both resonant and antiresonant processes. Evidently,
the treatment of these quantum processes is impossible within
conventional models based on classical (or perturbative)
descriptions of electron-field interaction (see, \textit{e. g.},
Refs. \cite{Galitskii,Goreslavskii}). Therefore, there is a
substantial improvement of the discussed diagrammatic approach to
describe the vacuum-induced processes in semiconductor structures
as compared to other methods used before.

\begin{figure}[t]
\includegraphics[width=.8\linewidth]{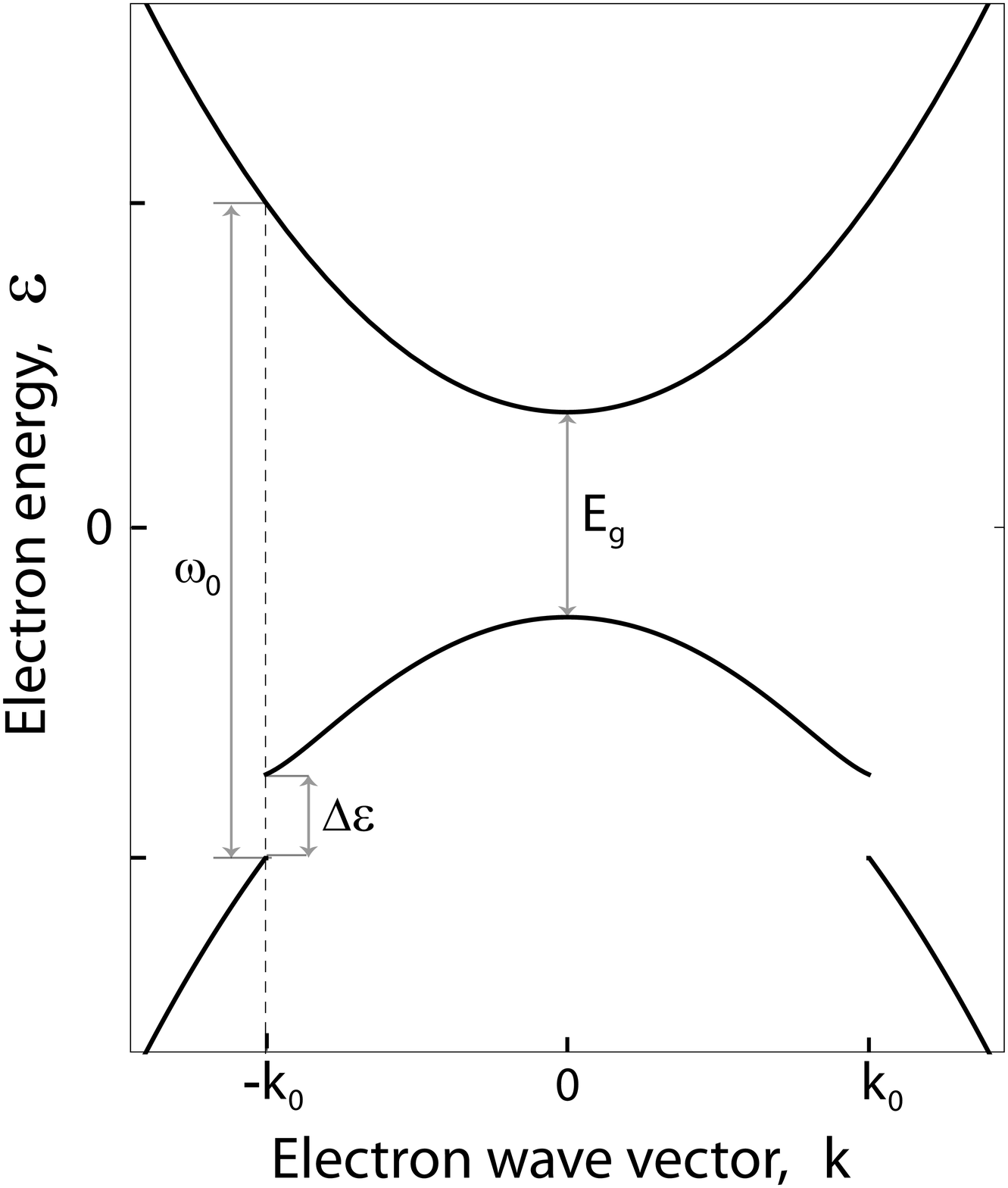}
\caption{(Color online) The gapped electron energy spectrum in a semiconductor induced by vacuum fluctuations of electromagnetic field in a one-dimensional microcavity.} \label{Fig3}
\end{figure}

\textit{Results and discussion.---}The energy spectrum of emergent
electron-photon quasiparticles, $\varepsilon(k)$, can be
obtained by finding poles of the electron Green's function given
by Eq.~(\ref{Greenf}). With this goal in mind, let us calculate
the self-energy operator $\Sigma_v(\varepsilon,k)$.
Substituting the bare Green's functions (\ref{DO}) and (\ref{G0c})
into Eq.~(\ref{Sigma}) and performing the integration over the
photon energy $\nu$, we find three poles located at energies
$\nu_1=\omega_0(q)-i\delta$,
$\nu_2=-\omega_0(q)+i\delta$, and
$\nu_3=\varepsilon-E_c(k-q)-i\delta$. The
contour integration can be made over the positive half-plane,
enclosing the pole $\nu_2$. Then, using Eq.~(\ref{oq}) and
applying the residue theorem, we can write Eq.~(\ref{Sigma}) as
\begin{equation}
\label{Sigma1}
\Sigma_v(\varepsilon,k)=\int_{-\infty}\limits^{\infty}\frac{g^{2}Ldq}{2\pi\left[\omega_0(q)+\varepsilon-E_{c}(k-q)-i2\delta\right]}.
\end{equation}
Performing the integration in Eq.~(\ref{Sigma1}), we arrive to the
expression
\begin{equation}
\label{Sigma3} \Sigma_v(\varepsilon,k)= \frac{\mu g^{2}
L}{\hbar^2\sqrt{\alpha(\varepsilon,k)}}\frac{\alpha(\varepsilon,k)}{\left|\alpha(\varepsilon,k)\right|},
\end{equation}
where
\begin{equation}
\label{alfa} \alpha(\varepsilon,k)=\frac{2\mu}{\hbar^2}\left[
\varepsilon+\omega_0-\frac{E_g}{2}-\frac{\hbar^2k^2}{2m}(1+\mu/m)
\right]
\end{equation}
and $\mu=m_0m_e(m_e-m_0)^{-1}$. The poles of the renormalized
Green's function (\ref{Greenf}) are given by
\begin{equation}\label{DE}
1-\Sigma_v(\varepsilon, k)G_v^0(\varepsilon,k)=0.
\end{equation}
Substituting Eqs.~(\ref{G0v}) and (\ref{Sigma3}) into
Eq.~(\ref{DE}) and solving this equation, we can find the energy
spectrum of electrons dressed by vacuum fluctuations,
$\varepsilon(k)$, which is shown schematically in
Fig.~\ref{Fig3}. The remarkable feature of this spectrum consists
in the vacuum-induced gap $\Delta\varepsilon$ which takes place at
the electron wave vector where the difference of energy between
the conduction band and the valence band is equal to the energy of the photons with $q=0$. It follows from Eq.~(\ref{DE}) that in the physically
relevant case $\mu\approx m_0$ this gap appears at the electron
wave vectors $k_0=\pm\sqrt{m_{e}(\omega_0 - E_g)}/\hbar$,
and its value is
\begin{equation}\label{Gap}
\Delta\varepsilon=\left(\frac{m_0^{1/2}g^2L}{\sqrt{2}\hbar}\right)^{2/3}.
\end{equation}
It should be noted that both the parameter describing the strength
of electron-photon interaction in the cavity,
$g^2L=|d_{cv}|\sqrt{{\varepsilon_0}/{2 \epsilon \epsilon_{0}S}}$,
and the gap Eq.~(\ref{Gap}) do not depend on the cavity length $L$ and, therefore, the developed theory is consistent in
the limit $L\rightarrow\infty$.

The calculation of the vacuum-induced gap (\ref{Gap}) for various semiconductor materials is presented
in Tab. 1. It can be seen from Table 1 that the bandgap can be macroscopically large.
Therefore, the discussed gap opening can be observable in the current experiments.
\begin{table}[h]
\begin{tabular}{|l|l|l|l|l|}
\hline
Semiconductor             & CdTe & ZnO  & GaN & GaAs \\ \hline
$L_0$ (nm)                & 224  & 132  & 100  & 160  \\ \hline
$\Delta\varepsilon$ (meV) & 0.03 & 0.04 & 0.06& 0.07 \\ \hline
\end{tabular}
\caption{Vacuum-induced bandgap $\Delta \varepsilon$ in different semiconductor materials and cavities.}
\end{table}

As we previously mentioned, the discussed effect strongly depends on the cavity geometry. Namely, the gap opening occurs only in a one-dimensional resonator where  photon cavity modes are confined along the $y,z$ axes (see Fig.~\ref{Fig1}). Physically, this follows from the singularity of the photon density of states, which takes place in one-dimensional photonic systems at the photon wave vector $q=0$. As a result of the singularity, interband
electron transitions induced by virtual photons with
$q=0$ (vertical transitions) prevail over other possible
transitions. Therefore, the gap is opened at the electron wave vector
$k_0$, where the interband distance is equal to the
energy $\omega(0)$ (see Fig.~\ref{Fig3}). On the contrary, in
conventional planar (2D) photonic cavities  there is no singularity in the photon density of states. Therefore
virtual photons with different wave vectors $q$ and energies $\omega(q)$ shift different electron energy
levels equivalently and the gap $\Delta\varepsilon$ is absent in this case.

In order to verify the developed diagrammatic approach, let us
apply it to a semiconductor system excited by a strong
laser-generated electromagnetic field. This system can be
described using the methods of classical electrodynamics, and the theory of
the gap opening induced by a laser has been elaborated in details
\cite{Galitskii,Goreslavskii,Kibis_PRL2011,Kibis_PRB2012}.
Therefore, it is instructive to compare this well-known theory
with results obtained from the discussed diagrammatic approach in
the limit of large photon occupation numbers. For the sake of
comparison, we will consider the physical situation where a cavity
is absent and a semiconductor structure is irradiated by a strong
laser beam. In contrast to the previously considered case of
vacuum fluctuations, the wave vector of laser-emitted photon
$q_l$ is fixed. To account for this fact in the
self-energy (\ref{Sigma}), we do not need to perform the integration over $q$, but account for the macroscopic population of the mode. The photon propagator under this assumptions is given by the expression
\begin{equation}
D^0(\nu)= \frac{2 N_{l} \omega_l}{(\nu- \omega_l + i\delta)(\nu+
\omega_l - i\delta)}, \label{D0_L}
\end{equation}
where $\omega_l$ is the energy of the laser-emitted photon and
$N_{l}$ is the photon occupation number of the laser mode.
Substituting the propagator (\ref{D0_L}) into Eq.~(\ref{Sigma})
and performing the integration over the photon energy $\nu$, we
arrive to the self-energy
\begin{equation}
\label{Sigma_L}
\Sigma_v(\varepsilon,k)=\frac{g^{2}N_{l}}{\omega_l +
\varepsilon - E_c(k)}.
\end{equation}
Solving Eq.~(\ref{DE}) with the usage of Eq.~(\ref{Sigma_L}), we
can write the energy spectrum of electrons dressed by the laser
field, $\varepsilon$, as
\begin{eqnarray}
\label{omegac_L} \varepsilon(k)&=&\frac{E_c(k)+
E_v(k)}{2}-\frac{\omega_l}{2}+\frac{k_l-k}{2|k_l-k|}\nonumber\\
&\times&
\sqrt{(\hbar\Omega_R)^2+(\omega_l-E_c(k)+E_v(k))^2},
\end{eqnarray}
where ${k}_l = \sqrt{m_{e}(\omega_l - E_g)}/\hbar$, and
$\Omega_{R} = 2g\sqrt{N_{l}}/\hbar$ is the Rabi frequency of
interband laser-induced electron transitions. The eigenenergies
(\ref{omegac_L}) describe the energy spectrum of electrons in the
valence band dressed by the laser field. It follows from
Eq.~(\ref{omegac_L}) that the field-induced gaps within the
valence band, $\Delta\varepsilon=\hbar\Omega_R$, take place at
electron wave vector $k_l = \pm\sqrt{m_{e}(\omega_l -
E_g)}/\hbar$ which satisfies the condition of interband electron
transitions, $E_c(k_l)-E_v(k_l)=\omega_l$. As
expected, this gapped spectrum exactly coincides with the energy
spectrum of dressed electrons obtained before from the classical
consideration of electromagnetic field \cite{Kibis_PRB2012}. This can serve as a confirmation for the consistency of the
developed theory.

\textit{Conclusions.---}We developed a non-perturbative diagrammatic approach based on the Green's function formalism,
which is applied to describe the renormalization of the electron energy spectrum of one-dimensional system consisting of quantum wire and cavity. The van Hove singularity in the photonic density of states enables a bandgap opening in the electron energy spectrum  induced by vacuum fluctuations in the cavity.  The approach is verified by comparison with well-known results obtained before within classical electrodynamics in the regime of strong electromagnetic field. The elaborated theory is of general character: it allows for unification of quantum and classical descriptions of the strong light-matter interaction in various semiconductor systems.

We thank T. C. H. Liew for useful comments and valuable discussions. The work was supported by Tier 1 project ``Polaritons for novel device applications,'' FP7 IRSES projects QOCaN and SPINMET, FP7 ITN NOTEDEV, the Russian Ministry of Education and Science, and the Russian Foundation for Basic Research (projects No. 13-02-90600 and 14-02-00033). T.~E.-O. and O.~V.~K. thank the University of Iceland for the hospitality. O.~K. acknowledges the support from the Eimskip Fund.


\end{document}